\documentclass[11pt,twoside]{article}

\usepackage{asp2004}

\begin{document}

\title{Polarimetry of Binary Stars}

\author{Nadine Manset}

\affil{Canada-France-Hawaii Telescope Corporation 
65-1238 Mamalahoa Hwy Kamuela HI 96743}

\begin{abstract}
Astronomical polarimetry is a powerful technique that can provide
physical information sometimes difficult or impossible to obtain by any
other type of observation. Almost every class of binary star can benefit
from polarimetric observations: pre-main-sequence objects, close or
contact binaries, mass-transfer systems, evolved binaries, cataclysmic
variables, eclipsing binaries, etc. In these systems, polarimetry can
help determine the geometry of the circumstellar or circumbinary matter
distribution, yield information on asymmetries and anisotropies,
identify obscured sources, map starspots, detect magnetic fields, and
establish orbital parameters, to name just a few examples. The orbital
inclination in particular is a very important piece of information for a
binary system because it can lead to the determination of the
components' masses, the fundamental parameter that determines a star's
initial structure and subsequent evolution. This review will illustrate
the usefulness of polarimetric techniques for the study of binary stars,
with examples of results obtained for a variety of binary systems, and
an overview of models, including those used to retrieve the orbital
inclination.
\end{abstract}

%%%%%%%%%%%%%%%%%%%%%%%%%%%%%%%%%%%%%%%%%%%%%%%%%%%%%%%%%%%%%%%%%%%%%%%%
%%%%%%%%%%%%%%%%%%%%%%%% INTRODUCTION %%%%%%%%%%%%%%%%%%%%%%%%%%%%%%%%%%
%%%%%%%%%%%%%%%%%%%%%%%%%%%%%%%%%%%%%%%%%%%%%%%%%%%%%%%%%%%%%%%%%%%%%%%%

\section{Introduction}
This review on polarimetry of binary stars has two main goals. First, to
demonstrate and illustrate the usefulness of polarimetry for the study
of binary stars, through various examples. As these examples will show,
there is quite an interesting variety of information that can be
gathered on binaries found all over the HR diagram. The second goal is
to present some of the models used to reproduce the observations and
extract information such as the orbital inclination. The sets of light
curves produced by these models help the interpretation of variations
seen in polarimetric observations.

\section{Polarization in Binary Stars}
The polarization of light observed in binary stars is not due to the
binarity itself but is produced by very general phenomena also seen in
single stars (and also in non-stellar objects): polarization by
scattering and by magnetic fields. What the binarity can potentially
introduce is variability of the polarization signal with
time. Otherwise, the polarization produced in binaries generally follows
well-known basic rules.

In the case of scattering (single or multiple), the strength of 
the polarized signal will depend on the nature of the scatterers
(electrons, molecules, dust grains), their characteristics (for dust
grains, composition and size), and their distribution (envelope, disk,
wind, stream, etc.). In the case of polarization produced by magnetic
fields, the signal will depend on the strength and orientation of the
field. In all cases, polarization also depends on the viewing angle.
This dependance on many and various parameters can be used to extract
information on the nature of scatterers, their spatial distribution, the
geometry of a system, magnetic field characteristics, orbital
parameters, etc. 

Because of symmetry, one spherical perfect star (perfect in the sense
that it has no spots, for example) does not produce polarization. In
1946, Chandrasekhar predicted that the {\it limb} of a star is polarized
at about 12\%, in the best case of a pure electron scattering
atmosphere. But if the star is spherical (symmetric) the {\it
integrated} polarization is zero. In a similar way, a spherical
distribution of scatterers will not produce polarization when the light
is integrated over the whole distribution. To produce polarization, one
has to break this symmetry.

For example, the surface of a spherical star can be deformed by
rotation. For binaries, tidal distortion will produce a similar
effect. In eclipsing binaries, the eclipse can reveal the limb
polarization, a phenomenon called the Chandrasekhar effect. For envelopes,
any non-spherical distribution can produce polarization: streams, disks,
aspherical winds, or spherical distributions with inhomogeneities, etc.

Almost any type of binary can produce observable
polarization. Pre-main-sequence stars have disks and envelopes where
scattering and polarization occur. Main sequence stars may have
shells and winds, and produce polarization in a similar way. Eclipsing
binaries can exhibit the Chandrasekhar effect. Cataclysmic variables and
other mass-transfer systems have streams of matter flowing from one star
to the other, with possibly accretion disks. Contact
binaries can have non-spherical surfaces. Mass-loss systems such as
Wolf-Rayet binaries can have aspherical winds. Envelopes can be
spherical but present inhomogeneities and unusual structures that will
produce a net polarization.

\section{Removal of the Interstellar Polarization}
In general, the observed polarization of any object is the sum of
intrinsic polarization produced at the source or in its immediate
vicinity, and interstellar (and sometimes intra-cluster) polarization
produced in the interstellar medium by grains located in the line of
sight. If the information sought is contained in the {\it
intrinsic} degree of polarization, the interstellar (IS) polarization
has to first be removed from the observations. There are two general
methods to do this.

One can use stellar polarization catalogs to estimate the strength and
orientation of the IS polarization in the vicinity of the
target. Compilations such as those of \citet{mathewson} and
\citet{heiles} are used to find the polarization of neighboring
stars within a certain angular distance of the target and, since the IS
polarization also depends on distance, at a similar distance. The
averaged polarization is assumed to describes the IS polarization for
the target and subtracted. The remaining polarization is only an estimate of
the intrinsic polarization. If the IS polarization in a region of the
sky is strong and uniform, its estimation will be easier to make and
more reliable. But if the IS polarization is weak or not very well
organized, the estimate will more uncertain, which will make the
estimate of the intrinsic polarization uncertain too. 

One can also use Serkowski's law for the IS polarization
\citep{serkowski}, assume that anything bell-shaped in the polarization
as a function of wavelength is of IS origin, and subtract it. Neither of
these method is perfect, though, and one should not over-interpret the
resulting calculated intrinsic polarization. For examples of IS
polarization subtraction, see \citet{paper3} or \citet{hoffman98}.

%%%%%%%%%%%%%%%%%%%%%%%%%%%%%%%%%%%%%%%%%%%%%%%%%%%%%%%%%%%%%%%%%%%%%%%%
%%%%%%%%%%%%%%%%%%%%%% 101 USES %%%%%%%%%%%%%%%%%%%%%%%%%%%%%%%%%%%%%%%%
%%%%%%%%%%%%%%%%%%%%%%%%%%%%%%%%%%%%%%%%%%%%%%%%%%%%%%%%%%%%%%%%%%%%%%%%

\section{Some of the 101 Uses of Polarimetry for the Study of Binary Stars}

%%%%%%%%%%% DISCOVERY

\subsection{Polarimetry as a Tool to Discover Binaries}
Polarimetry can be used to discover binary stars, like in the
serendipitous discovery made by \citet{gledhill}. The goal of their
project was to detect and image the dusty circumstellar (CS) envelopes
around protoplanetary nebulae using NIR polarimetric imaging. This
technique successfully separates the scattered light coming from the
faint envelope from the glare of the central object. As it can easily be
seen in the figures from that paper, the faint nebula is not seen in the
total flux images, whereas the polarized flux maps clearly show the
fainter nebulous structures.

The polarized flux contours usually reveal structures such as bipolar
nebulae, ellipsoidal envelopes, shells, and rings. But in one case
(IRAS~19475$+$3119) out of the 15 where a CS envelope was detected,
spiral arms are clearly seen. This is interpreted as being the signature
of an previously undetected binary star companion interacting with the
primary object. So in this first example, polarimetry has revealed the
presence of a binary star.

%%%%%%%%%%%%% CS ENVIRONMENT

\subsection{Polarimetry as a Probe of the Circumstellar or Circumbinary
Environment} 
When a binary (or a single star) is known or even just suspected of
having a disk or envelope, imaging polarimetry can be used to retrieve
information on the envelope itself or on its scatterers. For example,
\citet{potter2000} have obtained adaptive-optics, high-resolution
imaging polarimetry at 1.2$\mu$m of the circumbinary (CB) disk around
UY~Aur. The polarimetric observations show a centrosymmetric
pattern that is compared to a Mie scattering model. This analysis
suggests that the polarization comes mostly from the smallest grains
present in the disk ($\sim 0.03 \mu$m) and that the optical and IR light
comes from a large flattened disk. Indeed, a spherical distribution of
scatterers does not show changes in the the degree of polarization
measured as a function of position angle in the envelope. UY~Aur
presents a sinusoidal-like change in polarization as a function of
position angle, with maxima at the two 90$\deg$ scattering angles, the
distinguishing polarization signature of a disk.

A radial increase in polarization, of 10\% per arcsec, is also observed,
but it is unknown for now if it is a physical effect (multiple scattering
events) or an instrumental artefact (contamination from the unpolarized
PSF). 

Another CB disk studied in a similar way is GG~Tau (\citeauthor{silber}
\citeyear{silber}; \citeauthor{duchene} \citeyear{duchene}). Also see
these proceedings for Fukagawa et al.'s contribution for the binary
LkHa~198. The large scale vector pattern in this system is approximately
centrosymmetric around the primary star, indicating that the major
illuminating source in this region is the primary, not the companion
star. 

Other high-resolution, high-contrast IR observations of the CS
environment of (single) young stars were presented by Potter (these
proceedings) and can be used to get the physical parameters of the disks
such as the radial extent and inclination. Also of interest to this
subject are the reviews by M\'enard and by Tamura \& Fukagawa (these
proceedings). 

Finally, spectropolarimetry can also be used to determine the geometry and
temperature structure of circumstellar disks (see Bjorkman, Bjorkman, \&
Carciofi; Oudmaijer; Vink et al., these proceedings).  

%%%%%%%%%%%%%%% SF PROCESSES

\subsection{Polarimetry to Get Insight into Star Formation Processes}
Polarimetry can be used to study individual binaries, or a type of
binaries as a group, and even to get insight into physical processes,
for example, star formation processes.

Fragmentation (as in fragmentation of a cloud core or as in growth of an
instability at the outer edge of a disk) is now the best star formation
scenario. Fragmentation of cores in a cloud can create binaries with
non-coplanar orbits, but if stars grow from instabilities in the outer
parts of a disk, systems with parallel rotation axes will result.

To discriminate between the two scenarios, \citet{monin} used
polarimetry to estimate the relative orientation of the star plus disk
systems in pre-main-sequence wide binaries. Wide binaries were selected
so the polarization of each star plus disk could be measured separately.
The data consists of linear polarization measurements, with a careful
analysis to consider possible biases due to IS polarization. The result is
that for the systems where IS polarization does not introduce a bias, 4
systems have parallel axes, and one, GI Tau + GK Tau, clearly does
not.

Therefore, the alignment seen is thought to result from the initial
binary formation rather than from tidal interaction (which would produce
more non-parallel systems). This result was confirmed by
\citet{jensen}. For 8 out of 9 binary systems located in the Tau-Aur and
Sco-Oph star forming regions, the positions angles of each component of
a pair are within 30$\deg$ of one another, whereas the five triplet and
quadruple systems appear to have random axis orientation. See also the
contribution of Wolf et al. (in these proceedings) who found that disks
in binary systems are preferentially aligned.

So in these projects, the polarimetry has allowed conclusions that go
far beyond the properties of binary systems but include insight into
fundamental physical processes.

%%%%%%%%%%%%% REFLECTION

\subsection{Polarization by Reflection}
So far, the examples have dealt with scattering off particles located in
the stars' environment. Scattering can also occur at a star's surface,
as was first evidenced by \citet{rudy77}. The binary $\mu$ Herculis is
a close binary where the secondary fills its Roche lobe, and where there is
spectroscopic evidence for CS matter. 

The linear polarization data show low-amplitude second-harmonics
variations of 0.06\% amplitude in Stokes parameter $Q$. These periodic
variations can be caused by the reflection of light off the secondary
atmosphere, or by scattering off the stream of gas seen
spectroscopically. Modelling discriminates between the two possibilities
and favors the reflection mechanism.

There are not many cases of the reflection mechanism in the
literature. A more recent example can be found in \citet{berdyugin99}
where 0.3-0.4\% polarization variations are seen for LZ~Cep. In this
case, it is thought that the reflection mechanism is clearly the only
possible explanation because there is no spectroscopic evidence for CS
material, and the variations are too large to be due to tidal
deformation. In addition to pinpointing the polarization mechanism, the
authors were able to recover the orbital inclination.

%%%%%%%%%%%%%%% CHANDRASEKHAR

\subsection{The Chandrasekhar Effect in Eclipsing Binaries}
The limb polarization effect, or Chandrasekhar effect, was predicted in
1946, but only observed in 1983\footnote{In his summary presentation,
Roger Hildebrand told an interesting story about this. Chandrasekhar
learned almost simultaneously that he had won the Noble prize and that
the limb polarization effect had been detected. Apparently,
Chandrasekhar was more delighted with the observational proof of his
prediction rather than by receiving the Nobel prize!}
\citep{kemp83}. The data used for this discovery consisted in white light
observations of Algol taken mostly around primary eclipse. Even though
the amplitude of the polarimetric variations are only of the order of
0.01\%, primary eclipses clearly present a sharp positive peak with
weaker negative lobes in $Q$, and a S-shaped pattern of the $U$ Stokes
parameter. The patterns observed are compatible with the Chandrasekhar
effect. 

The Chandrasekhar effect was studied theoretically five years later by
\citet{landi} and analytical expressions were derived for the
polarization expected during an eclipse due to limb polarization
produced in a very thin annulus around the limb. In addition to a thin
annulus approximation, this study concerned detached binaries with
spherical stars, and no reflection effect. Using the best estimates by
\citet{kemp83} for the radius and eclipse geometry for Algol,
\citet{landi} were able to reproduce the general shape of the
polarimetric variations during the eclipse. 

Another 5 years later, \citet{wilson} made further improvements with a
more general binary star model that includes tidal and rotational
distortion, gravity and limb darkening, reflection effect, without the
thin ring approximation. The resulting $Q$ and $U$ polarimetric curves
show more structures. The scale of the variations, less than 0.01\% help
explain why this effect had not been observed before.

The eclipsing binary V444~Cyg, composed of a Wolf-Rayet and an O star,
was observed by \citet{robert90} and shows clear and phased-locked, and
good amplitude variations of about 0.3\%. The second harmonic
information was used to retrieve the orbital parameters, while
modeling the eclipse yielded information on the size of the stars. 
 
%%%%%%%%%% CIRCULAR

\subsection{Circular Polarization in Binaries}
The examples presented so far used linear polarimetry, but binaries with
strong magnetic fields, such as cataclysmic variables, present
significant levels of circular polarization. See in these proceedings
the contribution by Rodrigues et al.

A recent example of circular polarimetry of a polar (a sub-class of
cataclysmic variable in which the magnetic field of the white dwarf is
so strong that the white dwarf is synchronous with the secondary) can be
found in \citet{romero}. The linear polarization is relatively stable
but the circular polarization can be at times positive or negative. When
both signs of circular polarization are observed, the white dwarf is
accreting matter on two distinct spots. But spots can be transient, as
sometimes the circular polarization only shows one sign.  Models are
used to recover the orbital inclination and the latitude of the spots on
the stellar surface.

Other examples of circular polarization produced by cyclotron emission
regions can be found in \citet{schwope}, \citet{piirola},
\citet{katajainen}. 

%%%%%%%%%%%%%%%%%%%%%%%%%%%%%%%%%%%%%%%%%%%%%%%%%%%%%%%%%%%%%%%%%%%%%%%%
%%%%%%%%%%%%%%%%%%%%%%%%% BME %%%%%%%%%%%%%%%%%%%%%%%%%%%%%%%%%%%%%%%%%%
%%%%%%%%%%%%%%%%%%%%%%%%%%%%%%%%%%%%%%%%%%%%%%%%%%%%%%%%%%%%%%%%%%%%%%%%

\section{Orbital Inclination}

\subsection{The Basic Models}
A few theoretical models describe polarimetric variations of binary
systems and can usually be used to find the orbital inclination. Such
information is very valuable, since it can lead to the determination of
the absolute masses of the stars, the fundamental parameter for a
star. With the mass, one can determine the nature of a binary (for e.g.,
if it has a compact companion or something more massive; see
\citeauthor{dolan89} \citeyear{dolan89} or \citeauthor{dolan92}
\citeyear{dolan92}), one can make comparisons with theoretical models of
star formation or evolution, study the IMF, etc. 

\citet{rudy78} have shown that in the plane of the Stokes parameters $Q$
and $U$, the polarization as a function of time traces an ellipse, twice
per orbit, and the eccentricity of this ellipse depends only on the
orbital inclination. The inclination found is independent of size, shape
and position of the circumstellar region.

A very similar but more general and more often used method is that of
\citet{bme} (hereafter referred to as BME), who use a first and
second-order Fourier analysis of the Stokes curve to find, in addition
to the orbital inclination, moment integrals of the distribution.

The BME formalism includes the following assumptions:
\begin{itemize}
\item the binary star is in a circular orbit
\item the stars are point-like
\item photometric variability, if any, is low
\item the scatterers are electrons
\item the scatterers' distribution is arbitrary but co-rotating (the
electron density is fixed in a coordinate system in the which the stars
are fixed)
\item the material is optically thin so only single scattering is
considered 
\end{itemize}

Observations are represented as first and second harmonics of
$\lambda=2\pi\phi$, where $\phi$ is the orbital phase:\\
\begin{eqnarray}
Q &=& q_0 + q_1 \cos \lambda + q_2 \sin \lambda + q_3 \cos 2\lambda +
q_4 \sin 2\lambda, \label{eq-qfit}\\
U &=& u_0 + u_1 \cos \lambda + u_2 \sin \lambda + u_3 \cos 2\lambda +
u_4 \sin 2\lambda. \label{eq-ufit}
\end{eqnarray}

This representation is simply a low-order Fourier analysis and is very
general. But the coefficients of the fit can be used to get the
inclination, using the first (Equation \ref{EQ-iO1-p1}) or
second (Equation \ref{EQ-iO2-p1}) order Fourier
coefficients, although it is expected that second order variations will
dominate:\\
\begin{eqnarray}
\left[ \frac{1-\cos i}{1+\cos i} \right]^2 &=& \frac{(u_1+q_2)^2 +
(u_2-q_1)^2}{(u_2+q_1)^2 + (u_1-q_2)^2} \label{EQ-iO1-p1},\\
\left[ \frac{1-\cos i}{1+\cos i} \right]^4 &=& \frac{(u_3+q_4)^2 +
(u_4-q_3)^2}{(u_4+q_3)^2 + (u_3-q_4)^2} \label{EQ-iO2-p1}.
\end{eqnarray}

The BME formalism also returns the axis' orientation on the sky, four
integrals of the density distribution, the interstellar $U$ component
and the interstellar $Q$ component in combination with the fifth
integral over the envelope. Although the integrals of the density
distribution are also very interesting, their interpretation proves
difficult, and the BME formalism is mostly used to recover the orbital
inclination and axis orientation on the sky.

\subsection{Improvements to the BME Formalism}
The BME formalism has a specific set of assumptions which are not always
adequate to model a target or class of objects. Numerous improvements
and extensions to the BME model have been made, some of which will be
mentioned here. For additional references and information, the reader is
invited to consult the papers mentioned below.

\citet{brown82} have relaxed the assumption of corotation implicit in
the models developed up until then and studied the polarimetric
variations produced by a {\it localized} scattering region in an
eccentric orbit. Whereas circular orbits only produce second harmonics,
eccentric orbits add first and third harmonics. Moreover, erroneous
orbital parameters can be found if it is assumed that the orbit is
circular when it is in fact eccentric. \citet{simmons84} made similar
studies to apply to the X-ray transient AO~538-66, and also present a
correction to equations found in \citet{brown82}. An eccentric binary
model was also used by \citet{robert92} to study the Wolf-Rayet (WR)
binary EZ~CMa, also taking into account the fact that WR envelopes are
{\it extended} and not localized near one star.

Arbitrary scattering mechanisms (as long as they are spherically
symmetric, e.g., Mie scattering on dust grains) were studied by
\citet{simmons82,simmons83}, who re-derived the BME equations as a
special case.

% \citep{fox91} is for single stars - occultation and depolarization

\citet{fox94} has incorporated the {\it finite} size of stars in the BME
formalism, and found no change to the BME results when there is no
occultation of the scatterers, but additional Fourier harmonics when the
stars occult the scatterers.

See also \citet{bastien88} for additional details on the BME model and
results for early-type binaries, including the moment integrals over their
density distributions.

\subsection{Biases and Other Statistical Effects}
In their model, \citet{rudy78} use a second-order fit to find the
eccentricity of the $QU$ loop, which is related to the orbital
inclination $i$. The error on $i$ is then found by the propagation of
errors on the regression coefficients. It should be noted that this
formal error does not take into account the errors due to incorrect
modeling.

\citet{simmons80} have shown that taking only the formal
error, and ignoring the effects of noise and incorrect modeling can lead
to an underestimation of the confidence interval for $i$.  So they
propose an analytic method for finding  confidence intervals that
then give the range of $i$ over which the model provides an acceptable
fit to the data.

\citet{aspin81} have found that the inclination found depends on the
real inclination and on the observational errors. They derive the
standard deviation necessary, $\sigma_{nec}$, to determine the 
inclination to $\approx \pm 5 \deg$, with a 90\% confidence level. In
practice, these calculations give the lowest possible inclination that
can be reliably found according to the quality of the data at hand
(number of observations, observational errors, amplitude of the
polarimetric variations).  This method is useful to ascertain what can
be done with the data, but does not actually give, for a given set of
observations, a precision on the inclination found by the BME model, nor
does it tell if the inclination could still be found to, say,
$\pm10\deg$, or with a lower significance.

\citet{simmonsetal82} have shown that models find inclinations higher
than reality, and more so for noisy data or low inclinations. 

\citet{wolinski94} have also studied the confidence intervals for
orbital parameters determined polarimetrically. They used Monte Carlo
simulations to produce synthetic polarimetric curves to which Gaussian
noise was added. The result is analyzed with the BME model, and with the
use of graphs, one can estimate the confidence intervals for the orbital
inclination and other parameters returned by the BME model. With their
Fig.~5, one can find, for a given standard deviation over amplitude of
the variations ratio, the critical value of inclination below which the
$1 \sigma$ confidence interval extends to $i=0\deg$. This method is
again useful to ascertain what can be done with the data. Fig.~4 of that
same paper gives, for four levels of data quality, the $1 \sigma$ and $2
\sigma$ confidence intervals on the inclination found by the BME
model. These graphs can be used to read confidence intervals, although
interpolation between the curves is necessary to get confidence
intervals that go with the quality of the data at hand.

The BME formalism and its various extensions have been successfully used
to find the orbital inclination of WR binaries (see for example,
\citeauthor{drissen86a} \citeyear{drissen86a}, \citeyear{drissen86b}; 
\citeauthor{moffat90} \citeyear{moffat90}, \citeyear{moffat98};
\citeauthor{marchenko98} \citeyear{marchenko98}), 
massive interacting binaries (\citeauthor{berdyugin98a}
\citeyear{berdyugin98a}; \citeauthor{berdyugin98b}
\citeyear{berdyugin98b}), 
close binaries (e.g. \citeauthor{luna} \citeyear{luna}), 
X-ray binaries, etc. 

%%%%%%%%%%%%%%%%%%%%%%%%%%%%%%%%%%%%%%%%%%%%%%%%%%%%%%%%%%%%%%%%%%%%%%%%
%%%%%%%%%%%%%%%%%%%%%%% MODELS %%%%%%%%%%%%%%%%%%%%%%%%%%%%%%%%%%%%%%%%%
%%%%%%%%%%%%%%%%%%%%%%%%%%%%%%%%%%%%%%%%%%%%%%%%%%%%%%%%%%%%%%%%%%%%%%%%

\section{Models}
Models of the polarization and polarimetric variations generated in
binary stars fall under two broad categories: polarization produced by
scattering (mostly linear polarization), and cyclotron emission in star
spots (mostly circular polarization). Thomson or Mie single scattering
can be readily studied with simple formulas. Multiple scattering
requires more complex Monte Carlo simulations. 

Modeling can help see what are the effects of a specific geometry, 
scattering mechanism, eccentric orbits, eclipse geometry, etc. Modeling
can be also used to check if the BME formalism to recover the orbital
inclination still works for cases initially not considered. A few of
those models will be presented here.

\subsection{Polarimetry of Pre-Main-Sequence Binary Stars}
Observations of pre-main-sequence stars have exploded in the past two
decades, and it was only a question of time before somebody would try
the BME formalism on T~Tauri and Herbig~AeBe binaries to get the orbital
inclination, masses, and then compare observed masses to theoretical
models.

Unfortunately, PMS binaries do not fit the assumptions behind the BME
formalism: orbits of short-period spectroscopic binaries are
significantly non-circular (up to $e=0.6$), and the scatterers are dust
grains. \citeauthor{paper1} \citeyearpar{paper1,paper2} have
investigated the effects of Mie scattering and eccentric orbits on
polarimetric variations of binary stars and on the inclination found by
the BME formalism. The numerical simulations also include pre- and
post-scattering extinction factors which were not part of the BME
formalism. It is found that orbital eccentricity introduces first
harmonic variations, as had been found earlier. Asymmetric scattering
functions of dust grains can also introduce first harmonics in the
variations. More importantly, the BME equations can be used to get the
inclination. For low eccentricities, $e \la 0.3$, the
inclinations can be found with the first or second-order
coefficients. For the high eccentricities, $0.3 < e < 0.6$, only the
first-order coefficients should be used.

Those results were applied to one Herbig~AeBe binary \citep{paper3}, and
about two dozens T~Tauri binaries \citep{paper4,paper5}. The majority of
those binaries, 68\%, have intrinsic polarization above 0.5\% at
7660\AA. Many of those binaries do not present any evidence other than
polarimetric for the presence of dust, indicating that polarimetric
techniques are more sensitive to the presence of dust than photometric
or spectroscopic ones.  A handful of those binaries present interesting
periodic and phased-locked variations. Unfortunately, they also exhibit
stochastic polarimetric variations that, added to the relatively
low-amplitude periodic variations, prevent the BME formalism from
getting any sensible inclination.

\subsection{Self and Externally Illuminated Disks}
Self- and externally illuminated disks around one component of a close
detached binary are studied by \citet{hoffman} with a Monte Carlo
radiative transfer code that considers multiple Thomson scattering and
variable absorption. In a forthcoming paper (Hoffman, Nordsieck, \&
Whitney, in preparation), the problem is inverted and it is shown how an
analysis of polarimetric observations can provide information on the
geometrical and optical properties of circumstellar matter within a
binary system. Thus modelling goes further than obtaining orbital
inclination and can give insight into the characteristics of the disks
themselves. 

\subsection{Other Models}
\citet{dolan84} modeled the polarization produced when the light from
a Roche-lobe-filling primary in a close binary scatters off circumstellar
matter near the secondary. A regularized Monte Carlo approach allows to
include limb and gravity darkening caused by tidal distortion. The
resulting polarimetric variations have a morphology in agreement with the
theoretical predictions of the BME formalism. The inclination can also be
found using the BME equations even if this latter model includes several
simplifying assumptions.

Lucas, Fukagawa, \& Tamura also describe in these proceedings a
3-dimen\-sio\-nal Monte Carlo modelling of HL~Tau in the near infrared with
aligned non-spherical grains. Infrared linear polarimetry can provide
information on the magnetic field, the structure of the system and the
grain properties. Circular polarization models show that circular
polarization observations can help measure the magnetic field structure
in protostars.

A model of the accretion disk around the classical T~Tauri star AA~Tau
was presented by Pinte \& M\'enard (these proceedings). The quasi cyclic
variations of brightness and polarization, with a maximum polarization
when the system is faintest, were modeled using eclipses produced by
orbiting circumstellar material. The effects on the photometric and
polarimetric light curves of a warp at the inner edge of the accretion
disk and of hot spots on the stellar surface are studied through
multiple scattering Monte-Carlo simulations. The photopolarimetric
variations can be explained by the presence of a warp; hot spots
have a limited influence on them.

\section{Conclusions and Future Work}
Almost any type of binary system can benefit from polarimetric
observations. If there is scattering material (electrons, molecules,
dust grains) in the circumstellar or circumbinary environment, or if
magnetic fields are present, polarization of light is possible (if the
conditions, such as density of the scatterers or viewing angle, are
adequate). Scattering material is present in many binaries, from
pre-main-sequence systems to evolved mass transfer binaries, and magnetic
fields can be found in cataclysmic variables.

Polarimetry is a very powerful technique that can give information on
the geometry of a distribution (spherical envelope or flattened disk,
stream, inhomogeneities in a wind, etc.), the characteristics of
scatterers (density, nature, temperature, etc.), the location of spots
on a star's surface, and orbital parameters. It can also have a broader
application like the study of physical processes such as star formation.

Of high interest for binaries is getting the orbital inclination, since
it can lead to the determination of the absolute masses of the stars,
which is the fundamental parameter that governs a star's structure and
evolution. A few models such as the BME formalism exist to retrieve this
information from periodic and phased-locked polarimetric
variations. However, the results are affected by noise, amplitude of the
variations, and statistical biases. The method has nonetheless been
successfully used on various types of binaries, such as Wolf-Rayet
binaries, interacting pairs, eclipsing systems, etc.

In the era of 8- and 10-m telescopes equipped with sophisticated
instruments, one can easily foresee more high-resolution imaging that
takes advantage of adaptive optics, and observations of fainter targets
combined with improved uncertainties. Theorists will develop more models
or improve current ones to accommodate the numerous parameters found in
various binary systems; this will increase the diagnostic value of
polarimetric observations. Observers will greatly increase the value of
their polarimetric data if they use both linear and circular
polarimetry, at various wavelengths, and combine those polarimetric
observations with the other more commonly used techniques (imaging,
spectroscopy).

\end{document}